\pdfoutput=1
\documentclass[runningheads]{llncs}
\newif\ifExtended
\Extendedtrue
\usepackage[T1]{fontenc}
\usepackage[frozencache=true,cachedir=.]{minted}
\usepackage{graphicx}
\usepackage{mathpartir}
\usepackage{hyperref}
\usepackage{amsmath}
\usepackage{amssymb}
\usepackage{shuffle}
\usepackage{color}

\begin{document}
\ifExtended
\title{Actor Capabilities for Message Ordering (Extended Version)}
\fi
\author{Colin S. Gordon\inst{1}\orcidID{0000-0002-9012-4490} }
\authorrunning{C. S. Gordon}
\institute{Drexel University, Philadelphia PA 19104, USA
\email{csgordon@drexel.edu}}
\maketitle              %
\begin{abstract}
Actor systems are a flexible model of concurrent and distributed programming, 
which are efficiently implementable, and avoid many classic concurrency bugs 
by construction. However they must still deal with the challenge of 
messages arriving in unexpected orders.

We describe an approach to restricting the orders in which actors send messages to each other, 
by equipping actor references --- the handle used to address another actor ---
with a protocol restricting which message types can be sent to another actor and in which order using that particular actor reference.
This endows the actor references with the properties of (flow-sensitive) static capabilities, which we call actor capabilities.

By sending other actors only restricted actor references, actors may control which messages are
sent in which orders by other actors. Rules for duplicating (splitting) actor references ensure
that these restrictions apply even in the presence of delegation. 
The capabilities themselves restrict message send ordering, which may form the foundation for stronger 
forms of reasoning. We demonstrate this by layering an effect system over the base type system, 
where the relationships enforced between the actor capabilities and the effects of an actor’s 
behaviour ensure that an actor’s behaviour is always prepared to handle any message that may arrive.

\end{abstract}

\section{Introduction}
Actor systems have long been of interest since Hewitt’s original conception of the 
idea~\cite{hewitt1973universal}, 
both for anthropomorphic, intuitive appeal similar to that which boosted early interest in 
object-oriented programming (leading to work on active objects~\cite{nierstrasz1993regular,boer2017survey}) --- 
and because the programming model enjoys many natural benefits that make it particularly 
well-suited to structuring concurrent and distributed programs. 
Compared to popular shared-memory concurrency models like POSIX threads,
the actor model inverts control so that all 
actions in the system occur in response to asynchronous messages. This avoids classic issues with data 
races\footnote{Though actor implementations embedded in imperative programming languages may need additional 
techniques~\cite{gordon2012uniqueness,clebsch2015deny,milano2022flexible,swift6drf} to share 
data that was once or may again be mutable.} and deadlocks. 
Instead actor systems suffer from difficulties ensuring that messages are sent or received in expected orders~\cite{bagherzadeh2020actor,hedden2018comprehensive,torres2018study}, 
including the possibility that a message arrives before or after an actor is using a behaviour ready to process that 
particular message.

Static checking of message ordering in message passing systems is a classic problem, studied in many settings.
Nielson, Nielson, and Amtoft~\cite{amtoft1999,nielson1993cml} examined this problem for Concurrent ML (CML) 
in the 1990s, proposing what we now recognize as sequential~\cite{tate13} (a.k.a., flow-sensitive) type-and-effect systems 
for ensuring sends and receives in that synchronous message passing environment were in agreement between 
different threads. Work on session types began around the same time~\cite{Honda2008,honda1998language},
 initially for pairs of processes, and later for arbitrary numbers of processes.
 (And as it turns out, those can be formulated as effect systems as well~\cite{orchard2016effects}.) 
Each of these has a language for describing \emph{all} communication behaviour of a process, yielding
possibly-complex specifications. (We discuss this further in Section \ref{sec:relwork}.)

Effect systems are closely related to static \emph{capability systems},
which typically trade off some expressivity and precision for dramatic increases in brevity of specifications~\cite{gordon2020designing}.
Rather than allowing a program to do whatever it pleases 
and attempting to analyze that after the fact with effects, these systems require a specific capability for the program to perform any 
operation of interest. 
Dynamic capability systems do this via the dynamic semantics of the language: 
the operation literally cannot execute without the availability of a capability value~\cite{levy1984capability,miller2006robust}.
Actor systems are already a form of dynamic capability system --- an actor X can only send messages to another actor Y if it possesses the actor reference for Y, and provides it to the send operation.  Static capability systems instead use static checking — usually a type(-and-effect) system — to restrict what a region of code can do by restricting which (static) capabilities are accessible to that code. In this way, the set of input capabilities can give an upper bound on the effect of a program fragment~\cite{craig2018,gordon2020designing}.

A less widely-acknowledged aspect of static capability systems is that by restricting
the creation or duplication of capabilities, it becomes possible to enforce \emph{global} 
invariants with \emph{local} checks~\cite{gordon2020designing}, which has been employed 
throughout decades of work on what are now collectively known as \emph{reference capabilities}, 
where the capability restrictions are attached to each reference to an object. 
This encompasses many ownership (and universe) type systems~\cite{boyapati01,boyapati02,dietl07}, read-only 
reference systems~\cite{tschantz05,zibin07,zibin10,gordon2012uniqueness,giannini2019flexible}, and rely-guarantee reasoning applied between references~\cite{gordon2013rely,militao2014rely,militao2016composing,gordon2017verifying}, to name just a few variants.

In this paper we explore this idea adapted to \emph{actor references} rather than heap references, 
in a version of these ideas we call \emph{actor capabilities}. 
Each actor reference carries a restriction, given as a formal language over message types, restricting the order in which messages
may be sent using that particular reference; different references to the same actor may carry different capabilities/restrictions,
allowing highly \emph{localized} checks for message ordering.
We also make an unusual choice compared to most capability systems: rather than assuming 
each actor is created with a specific fixed language, we allow each actor to dynamically create \emph{new} capabilities for itself --- tracked
with an effect system. The result is a type system which combines flow-sensitive capabilities
with flow-sensitive (sequential~\cite{tate13}) effects in a highly-integrated way.

\section{An Intuitive Explanation of Actor Capabilities}
\label{sec:intuition}
Current actor system implementations come in two varieties: those which allow actors to send
any message type to any actor at any time, and those which attach a type parameter to actor references (handles
used to address another actor) which restrict the types of messages that can be sent.
The former category includes early versions of Akka and modern versions of Akka.NET,
as well as all dynamically-typed actor languages and libraries (Erlang, Elixir, etc.).
The latter category includes modern Akka, which we focus on because it is closest to what we propose.
\emph{All} actor systems in either category effectively handle messages by case analysis of the message's dynamic type.
Akka (now) uses the Scala type \mintinline{Scala}{ActorRef[T]} to represent a handle to send a particular
actor messages of type \mintinline{Scala}{T}. This is both a dynamic capability and a static capability:
\begin{itemize}
\item It is impossible to send an actor a message without obtaining an \mintinline{Scala}{ActorRef[T]} referring to it.
      These actor references cannot be forged by a program. Thus an actor reference is effectively a form of dynamic
      object capability~\cite{miller2006robust}.\footnote{We are intentionally side-stepping discussions of
      capability-safety and ambient authority raised by the ability to look up an actor reference by the actor's name.}
\item To send an actor a message of type \mintinline{Scala}{U}, code must use an \mintinline{Scala}{ActorRef[T]} where \mintinline{Scala}{U <: T}
      (\mintinline{Scala}{U} is a subtype of \mintinline{Scala}{T}).  To send some other message of type \mintinline{Scala}{B},
      the code would still need an \mintinline{Scala}{ActorRef[B]} --- even if a developer knew (correctly!) that the recipient
      would \emph{always} accept a \mintinline{Scala}{B}, the corresponding actor reference would be required by the type system.
      Multiple actor references to the same actor with different (incomparable!) message type bounds can exist. Thus this
      already has much in common with reference capability systems.
\end{itemize}
Akka also ensures that if an \mintinline{Scala}{ActorRef[T]}  exists for a specific actor, then that actor will \emph{always}
accept messages of type \mintinline{Scala}{T}. With the capability properties above, this invariant ensures that all messages
that can be sent will be handled.
The code of Figure \ref{fig:typedakka} offers a useful starting point to dig into this a bit more,
to call out the specific aspects of this situation we must generalize to deal with the set of valid messages changing over time.

\begin{figure}[t]\scriptsize
\begin{minted}{Scala}
import akka.actor.typed.*
import akka.actor.typed.scaladsl.*
object ExampleActor {
  def apply(): Behavior[MessageBase] = 
    Behaviors.setup((context: ActorContext[MessageBase]) => {
      lazy val beh : Behavior[MessageBase] = // Recursively-defined behaviour
        Behaviors.receiveMessage((msg: MessageBase) => {
          msg match { // Inspect the message type
            case Heartbeat(sender) => // if it is a heartbeat...
              val returnAddress : ActorRef[Heartbeat] = context.self
              sender ! HeartbeatResponse(returnAddress) // reply
              beh // Return next behaviour from this case
            ... // other message cases omitted
          } })
      beh // Make beh the initial behaviour
    }))
}
\end{minted}
\vspace{-2em}
\caption{A simple Scala program using the Typed Akka API}
\label{fig:typedakka}
\end{figure}

The \mintinline{Scala}{ExampleActor} is an object which when invoked (i.e., by \mintinline{Scala}{ExampleActor()} implicitly
calling the \mintinline{Scala}{apply} method) creates a new behaviour, defined in the presence of an
Akka-selected \mintinline{Scala}{ActorContext} for the newly-created actor; for our purposes, this is effectively a container for
the new actor's self reference \mintinline{Scala}{context.self} of type \mintinline{Scala}{ActorRef[MessageBase]}
(though of course Akka uses it for much more). The new actor is created by \mintinline{Scala}{Behaviors.setup}.
\mintinline{Scala}{beh} is a recursively defined \mintinline{Scala}{Behavior[MessageBase]}, indicating that this is a behavior which accepts
messages of type \mintinline{Scala}{MessageBase} (not shown).
Akka's functional API (the one used in this example\footnote{Akka also has an object-oriented API, and both are heavily used,
but the functional API is quite close to both classic actor calculi~\cite{agha1986actors} and our formal development.})
requires each behaviour to return a new behavior to handle the next message, which
this code does in each branch of the \mintinline{Scala}{match} statement, returning itself (so after handling one message, the behavior
explicitly signals that it will be used to handle the next message as well).

The one shown case of the behavior responds to the \mintinline{Scala}{Heartbeat} subclass of \mintinline{Scala}{MessageBase}
by first extracting a new reference for the sender to directly message this actor in the future (the message may have been routed
indirectly through other actors). The type of the ``return address'' reference is specialized to its intended use: \mintinline{Scala}{ActorRef[Heartbeat]}.
\mintinline{Scala}{ActorRef[MessageBase]} is a subtype of \mintinline{Scala}{ActorRef[Heartbeat]} --- \mintinline{Scala}{ActorRef} is contravariant in its type parameter.
So via subtyping the \mintinline{Scala}{context.self} reference,
which can be used to send any subtype of \mintinline{Scala}{MessageBase},
can be coerced to a more restrictive type, which can only be used to send further \mintinline{Scala}{Heartbeat} messages.
This more restrictive reference is then sent back to the sender.

\subsection{A Capability View of Safety Guarantees}
Notice that we derive important safety guarantees from the type system:
{the original requestor can never use the \mintinline{Scala}{ActorRef} sent back with the response to send the \mintinline{Scala}{ExampleActor}
a message type it does not handle, and the actor is statically known to handle all messages it receives}.\footnote{We are using handle in a loose sense here;
the behaviour might simply ignore some subtypes of \mintinline{Scala}{MessageBase} (intentionally or due to a developer's mistake).}
We obtain this from the (sometimes implicit) three-way agreement across:
\begin{itemize}
\item The \mintinline{Scala}{ActorRef[MessageBase]} (from \mintinline{Scala}{context.self}) specifies the expected type of messages sent to the actor.
      In particular, all other actor references to this actor are derived from \mintinline{Scala}{context.self} or the spawn operation (which
      yields an \mintinline{Scala}{ActorRef} of the same type), and by contravariance specify only \emph{subtypes} of \mintinline{Scala}{ActorRef[MessageBase]} ---
      for any \mintinline{Scala}{ActorRef[T]} referring to this actor, \mintinline{Scala}{T <: MessageBase}.
\item The \mintinline{Scala}{Behavior[MessageBase]} specifies the expected message (super)type for incoming messages,
      and \mintinline{Scala}{context.self}'s type parameter to \mintinline{Scala}{ActorRef} will always match
      the behaviour's type parameter.
\item The send operation (\texttt{!} operator) at the recipient of the response (\mintinline{Scala}{sender})
      only allows messages of type \mintinline{Scala}{Heartbeat} to be sent via an \mintinline{Scala}{ActorRef[Heartbeat]}
\end{itemize}

\mintinline{Scala}{ActorRef[T]} can be seen as a family of static capabilities. Owning one for a particular \mintinline{Scala}{T}
\emph{only} permits sending values of (subtypes of) type \mintinline{Scala}{T}, even if the destination actor can accept additional
message types. The fact that each new capability to a certain actor is created with \emph{at most} the permissions of the capability it
was copied from means that creation of new capabilities / actor references collectively preserve a global invariant that no part of the
program possesses a capability to send anything but a \mintinline{Scala}{MessageBase} in this example.
This property of creating new \mintinline{Scala}{ActorRef}s from old is similar to a cross-cutting theme in much of the static
capability literature, though only some work makes this fully explicit~\cite{CastegrenW17,gordon2013rely,gordon2017verifying,militao2014rely,militao2016composing}
(as we will need to do later).
As all actor references are either returned from the actor creation (not shown) or derived from the actor's self-reference (\mintinline{Scala}{context.self}),
both of which have an \mintinline{Scala}{ActorRef} parameter matching the \mintinline{Scala}{Behavior} parameter, this collectively maintains the 
global property~\cite{gordon2020designing} that no actor reference can be used to send a message that will not be handled by the installed behaviour.
While each behaviour returns a possibly-different behaviour for handling future incoming messages,
the type system ensures the new behaviour's message type matches the current behaviour,
so previously-shared references to the
actor never become invalid (i.e., never permit sending messages unexpected by some future behaviour of the same target actor).
\looseness=-1

This is a valuable property, and the reason that newer versions of Akka switched \emph{to} this typed API, rather than using a general
\mintinline{Scala}{ActorRef} type without a type parameter (called the ``classic'' API), which could be sent messages of any type. However, it does come at a cost.
It is common for actors to move through multiple lifecycle phases, where some messages are (temporarily or permanently) irrelevant, yet
a correct program must still explicitly include code to handle those unexpected messages.
The older untyped Akka API allowed code to simply switch to a new behaviour which might not handle \emph{any} of the message types it previously
received! This was convenient for code that was correct, 
but nothing statically prevented bugs where one actor would send another a message
of a type it either expected to never see again or to not see yet.
Technically, the same can still happen with the typed API if \mintinline{Scala}{MessageBase} is actually \mintinline{Scala}{Object},
it is simply less pronounced because developers are encouraged to use smaller type hierarchies.
It would be desirable to preserve the kinds of safety guarantees targeted by this implicitly capability-based API,
while extending them to \emph{fully} resolve the question of certain message types arriving at unexpected times.

\subsection{Extending the Capability View for Message Ordering}
\label{sec:extending}
This capability-based view of how Akka (and similar actor systems) ensure all messages are handled can be extended to deal with
the case where actors' set of accepted messages change arbitrarily over time.

The key idea we will employ is to replace the single type parameter of Akka's \mintinline{Scala}{ActorRef} with a formal language
over types --- thus specifying a set of allowable \emph{sequences} of messages that may be sent to that actor. This allows the
specification of actor references that permit sending just a single message, a fixed number of messages,
sending a certain message type just once and others an unbounded number of times,
alternating message types in a particular way, or other interesting restrictions --- the notion of a formal language is quite general.

On top of this, we must preserve and extend the three-way agreement described above. When an actor reference is duplicated, we cannot (in general)
allow the reference to be duplicated at the same type. Otherwise an \mintinline{Scala}{ActorRef} that permitted sending exactly one message of type \mintinline{Scala}{T}
could first be copied, then used to send a \mintinline{Scala}{T}, then the copy could be used to send a \mintinline{Scala}{T}, circumventing the intended restriction
of the original capability. So we must adapt machinery for carefully tracking splitting of capabilities~\cite{gordon2013rely,gordon2017verifying,militao2014rely,militao2016composing,CastegrenW16,CastegrenW17} to ensure the number and relative ordering
of message sends are preserved when actor references are duplicated --- including handling the subtleties from the new copies being usable in any (possibly-interleaved) order.
Fortunately there are existing tools in formal language theory we can use to guide the design here.

The other pieces of preserving and extending this three-way correspondence are the send operation, and agreement between actor references and
behaviour typing. The send operation
clearly must be typed in a flow-sensitive manner, to track when a send ``uses up'' one allowed message type. The more complex piece is maintaining
agreement between these consumable actor reference capabilities and a more flexible behaviour typing that allows actors to start and stop
accepting messages of various types. The challenge is to ensure that previously-shared capabilities continue to remain valid
when an actor changes behaviours.
\looseness=-1

If we treat Akka's existing \mintinline{Scala}{ActorRef[T]} as a capability to send a sequence of messages of types in $\{\tau \mid \tau <: \texttt{T}\}^*$,
then ensuring every future handler for the actor must handle the same types is sufficient to ensure old capabilities remain valid, because the message
types are fixed permanently (invariant), so ensuring that the initial actor references and initial behaviour agree on \mintinline{Scala}{T}
at actor creation time is a sufficient check. 
But if both the permissions of remaining capabilities, and the future behaviours' domains change over time, we will
need additional machinery to synchronize those changes. We will use an effect system for this, where duplicating/splitting a capability
for the current actor will create obligations for future message handling tracked in effects.

\section{An Actor Capability Language}

\begin{figure}[t!]\scriptsize
\begin{mathpar}
    \begin{array}{rrcl}
    \textsf{Expressions} & e & ::= & n \mid b \mid \langle e,e\rangle \mid e \oplus e \mid \neg e \mid \mathsf{if}\,e\,e\,e \mid (\lambda_f x\ldotp e) \mid \rho \mid \mathsf{beh}_L(\overline{\tau\leadsto e}) \mid \mathsf{spawn}(e)\\
    \textsf{Operators} & \oplus & \in & + \mid - \mid \times \mid \div \mid \vee \mid \wedge\\
    \textsf{Actor Stores} & \sigma&\in & \mathsf{Location}\rightharpoonup \mathsf{beh}_L(\overline{\tau\leadsto e})\\
    \textsf{Message Queues} & \mu & \in & \mathsf{list}(\mathsf{Value}\times\mathsf{Type}\times\mathsf{Location})\\
    \textsf{Global Message Queues} & \gamma & \in & \mathsf{Location}\times\mathsf{Location}\rightharpoonup \mathsf{list}(\mathsf{Value}\times\mathsf{Type})
    \end{array}
    \\
    \fbox{$\mu;\sigma;e\xrightarrow{L}_\ell \mu';\sigma';e'$}
    \and
    \textrm{Booleans, natural numbers, pairs, and (recursive) functions omitted}
    \and
    \inferrule*[left=\scriptsize E-Self]{ }{ 
        \mu;\sigma;\mathsf{self}_L \xrightarrow{\{L\}}_\ell \mu;\sigma;\ell}
    \and
    \inferrule*[left=\scriptsize E-Spawn]{
        \mathsf{IsBehaviour}(v)\\
        \mathsf{fresh}\;\ell'
    }{
        \mu;\sigma;\mathsf{spawn}(v) \xrightarrow{\{\epsilon\}}_\ell \mu;\sigma\uplus{\ell\mapsto v};\ell'
    }
    \and
    \inferrule*[left=\scriptsize E-Send]{
        l\Downarrow_\ell \ell_\mathsf{dst};\mu;\sigma \\
        e\Downarrow_\ell v;\mu';\sigma'
    }{
        \mu;\sigma;\mathsf{send}_\tau(\ell_\mathsf{dst},v)\xrightarrow{\{\epsilon\}}_\ell \mu+(v,\tau,\ell_\mathsf{dst});\sigma;()
    }
    \\
    \fbox{$\sigma;\gamma\Rightarrow\sigma';\gamma'$}
    \and
    \inferrule[\scriptsize E-MsgHandled]{
        \tau\in\overline{\tau}\\
        \emptyset;\emptyset;e_\tau[\mathsf{msg}\mapsto v]\xrightarrow{L}^*_\ell \mu';\sigma';v'
    }{
        \sigma[\ell\mapsto \mathsf{beh}(\overline{\tau\leadsto e})];\gamma[\ell,\mathit{sender}\mapsto (v,\tau)::\overline{vs}]\Rightarrow \sigma[\ell\mapsto v']\uplus\sigma';\mathsf{enqueue}(\ell,\gamma,\mu)
    }
\end{mathpar}
where $\mathsf{enqueue}(\ell,\gamma,\mu)$ extends each $\gamma(\ell_\mathsf{destination},\ell)$ with $\mu(\ell)$ by appending, thus preserving message ordering between endpoints,
without committing to delivery order across actors.
\caption{An actor language, partial operational semantics}
\label{fig:lang}
\end{figure}
Figure \ref{fig:lang} gives the syntax, and dynamic semantics of a simple actor language (abbreviated to omit entirely standard aspects).
As in the most classic versions of semantics for the actor model~\cite{agha1986actors}, an actor is a reactive program,
executing only in response to an incoming message. The actor's message handling is defined by a \emph{behaviour}, which we model as a partial function from types to handling expressions for each accepted incoming message type.
In response to a message, the message handler may create other actors and/or send messages to existing (or newly-allocated) 
actors. 
$\mu;\sigma;e\xrightarrow{L}_\ell \mu';\sigma';e'$ says that with initial outgoing message queue $\mu$ and locally-spawned actors $\sigma$,
expression $e$ running as actor $\ell$ steps to $e'$, resulting in message queue $\mu'$ (which always extends $\mu$) and locally-spawned actors $\sigma'$ (which always extends $\sigma$).
The $L$ indicates the cumulative self-capabilities generated by the reduction, which is the empty string except when
reducing an annotated $\mathsf{self}$ expression.
Iterated single stepping shuffles these formal languages together --- an operation discussed in detail with the type system.
The language $L$ is purely an instrumentation and does not influence execution.
Local actor reduction is deterministic in this calculus.

The full program steps by reducing an individual actor's reaction to any message ready for delivery:
$\sigma;\gamma\Rightarrow\sigma';\gamma'$ has one rule, delivering an available message to an actor \emph{which is able to handle it} in the sense
of running a matching handler to completion (i.e., termination).
The handler body is expected to reduce to another behaviour, which may be the same behaviour 
or a new behaviour.
(Note that this is also how many modern actor frameworks, such as Akka, work as well.)
Message delivery is tracked per send-receive pair and is in-order between any pair of actors in a single direction, 
but incoming messages from different actors may be interleaved arbitrarily.
Configurations where all actors either diverge\footnote{
    Non-terminating message handlers are always a bug in actor programs, as they prevent the actor
    from handling any messages in the future. Another static analysis could be used to
    ensure termination.
}
for any of the next incoming messages or do not handle them are stuck.
Configurations where at least one actor may continue executing are \emph{not} stuck.
\looseness=-1

\paragraph{Failures}
In typical actor implementations, if an actor does not support a particular message type, 
either the message is dropped (the runtime simply doesn't deliver the message at all),
explicitly ignored / passes through a generic handler which matches on message type,
or an exception is delivered to the receiving actor.
Regardless of typical handling in implementations, sending a message to an actor that does not expect that message type
is typically an error we would like to statically prevent.
We abstract the three possibilities above as a stuck transition in our formalization.

\paragraph{Actor State and Recursive Behaviours}
In our calculus, there is no mutable state, so changes to the state of an actor between messages must be encoded by 
returning a new behaviour which somehow captures the updated state. Because our functions are always recursively defined,
we can for example encode a count of messages received as
\begin{equation}
    (\lambda_\mathit{f}\mathit{state}\ldotp \mathsf{beh}(\mathbf{0}\leadsto f\;(\mathit{state}+1)))\;0
    \label{eq:state_passing_example}
\end{equation}
That is, \emph{state} binds the current count, and in response to receiving a unit message, the new behaviour is
defined by recursively calling the function with the count incremented by one.
The issues involved in direct mutable state are discussed in Section \ref{sec:relwork}.

Note that \ref{eq:state_passing_example} above also highlights the construction of a recursive behaviour \emph{indirectly}
through the use of a recursive function. Real actor framework implementations frequently include direct self-references
in their behaviours (e.g., if a behaviour is modeled as an object, the standard \mintinline{Scala}{this} refers to the behaviour)
or have ways to implicitly repeat a behaviour if a change in behaviour is not explicitly requested.
Both of these can be encoded in our calculus.

\paragraph{Actor Self-Reference and Recursive Behaviours}
Actors may retrieve their own address using the \textsf{self} operator. Note that in our calculus,
\textsf{self} is an \emph{operator}, not a variable bound to the current actor's reference.
While \textsf{self} could be modeled as a variable, our type system treats \textsf{self} somewhat differently
from variables, for reasons explained in Section \ref{sec:effects}.

\section{Actor Capabilities For Guaranteed Message Handling}
\label{sec:typing}
\begin{figure}[t!]\scriptsize
\begin{mathpar}
    \begin{array}{rrcl}
    \textsf{Types} & \tau & ::= & \mathbb{B}\mid \mathbb{N} \mid \mathbf{1} \mid \tau\times\tau \mid \tau\xrightarrow{\chi}\tau \mid \mathsf{ActorRef}(L) \mid \mathsf{Beh}(L) \\
    \textsf{Effects} & \chi & \in & \mathcal{L}(\tau) \\ %
    \textsf{(Formal) Languages} & L & \in & \mathcal{L}(\tau)\\
    \textsf{Paths} & p & ::= & x \mid p.1 \mid p.2
    \end{array}
    \\
    \fbox{$\Gamma\vdash e : \tau \dashv \Gamma' \mid {\chi}$}
    \and
    \textrm{Rules for booleans, conditionals, numbers, and pairs omitted}
    \and
    \inferrule[\scriptsize T-App]{
        \Gamma\vdash e_1 : \tau\xrightarrow{\chi}\tau'\dashv \Gamma' \mid \chi_1\\
        \Gamma'\vdash e_2 : \tau\dashv \Gamma'' \mid \chi_2
    }{
        \Gamma\vdash e_1\;e_2 : \tau'\dashv \Gamma'' \mid \chi_1\shuffle\chi_2\shuffle\chi
    }
    \quad
    \inferrule[\scriptsize T-Lam]{
        \Gamma\vdash \Gamma \prec \Gamma\divideontimes\Gamma\\\\
        \Gamma,f:\tau\xrightarrow{\chi}\tau',x:\tau\vdash e : \tau'\dashv\Gamma\mid\chi
    }{
        \Gamma\vdash (\lambda_f x\ldotp e) : \tau\xrightarrow{\chi}\tau'\dashv \Gamma\mid\{\epsilon\}
    } 
    \and
    \inferrule[\scriptsize T-Contract]{
        \Gamma\vdash\tau\prec\tau'\divideontimes\tau''\\
        \Gamma,x:\tau',x:\tau'',\Gamma'\vdash e : \tau\dashv \Gamma'\mid\chi
    }{
        \Gamma,x:\tau,\Gamma'\vdash e : \tau\dashv \Gamma'\mid\chi
    }
    \and
    \inferrule[\scriptsize T-Weaken]{
        \Gamma,\Gamma'\vdash e : \tau\dashv \Gamma''\mid\chi
    }{
        \Gamma,x:\tau,\Gamma'\vdash e : \tau\dashv \Gamma''\mid\chi
    }
    \and
    \inferrule*[left=\scriptsize T-Var]{ }{\Gamma,x:\tau\vdash x:\tau\dashv\Gamma\mid \{\epsilon\}}
    \and
    \inferrule*[left=\scriptsize T-Self]{ }{\Gamma\vdash \mathsf{self}_L:\mathsf{ActorRef}(L)\dashv\Gamma \mid L}
    \and
    \inferrule*[left=\scriptsize T-Send]{
        \Gamma\vdash e : \tau\dashv \Gamma' \mid \chi\\
        \Gamma'\vdash_\mathsf{path} p : \mathsf{ActorRef}(L) \leadsto \mathsf{ActorRef}(\tau^{-1}L) \vdash \Gamma''\\
        \tau^{-1}L\neq\emptyset
    }{
        \Gamma\vdash \mathsf{send}_\tau(p, e) : \mathbf{0}\dashv \Gamma''\mid\chi
    }
    \and
    \inferrule[\scriptsize T-Spawn]{
        \Gamma\vdash e : \mathsf{Beh}(L) \dashv \Gamma' \mid \chi
        \\
        L' \subseteq L
    }{
        \Gamma\vdash \mathsf{spawn}(e) : \mathsf{ActorRef}(L') \dashv \Gamma' \mid \chi
    }
    \and
    \inferrule[\scriptsize T-If]{
        \Gamma\vdash e_c : \mathbf{B} \dashv \Gamma' \mid L_c\\
        \forall b\in\{t,f\}\ldotp \Gamma'\vdash e_b : \tau \dashv \Gamma_b\mid L_b
    }{
        \Gamma\vdash \mathsf{if}~e_c~e_t~e_f : \tau \dashv \Gamma_t\cap\Gamma_f \mid L_c\shuffle(L_t\cup L_f)
    }
    \and
    \inferrule*[left=\scriptsize T-Beh]{
      \forall i\in 1..n\ldotp 
        \Gamma,\mathsf{msg}:\tau_i \vdash e_i : \mathsf{Beh}\left((\tau_i^{-1}L)\shuffle L_i'\right) \dashv \Gamma_i'\mid L_i'
      \\
      \mathsf{disjoint}(\overline{\tau})  
    }{
      \Gamma\vdash\mathsf{beh}_L(\overline{\tau\leadsto e}^n) : \mathsf{Beh}(L) \dashv \epsilon \mid \{\epsilon\}
    }

    \\
    \fbox{$\Gamma\vdash \tau \prec \tau \divideontimes\tau$}
    \and
    \inferrule*[left=\scriptsize S-ActorRef]{ L' \shuffle L'' \subseteq L }{\Gamma\vdash \mathsf{ActorRef}(L)\prec \mathsf{ActorRef}(L')\divideontimes \mathsf{ActorRef}(L'')}
    \and
    \inferrule*[left=\scriptsize S-SelfSplit]{\tau\in\{\mathbb{B},\mathbb{N},\mathbf{1}\}\vee\tau=\tau'\xrightarrow{\chi}\tau''}{\Gamma\vdash\tau\prec\tau\divideontimes\tau}
    \;\;
    \inferrule*[left=\scriptsize S-Pair]{
        \Gamma\vdash \tau\prec\tau\divideontimes\tau\\
        \Gamma\vdash \sigma\prec\sigma\divideontimes\sigma\\
    }{
        \Gamma\vdash \tau\times\sigma\prec \tau\times\sigma \divideontimes \tau\times\sigma
    }
\end{mathpar}
\vspace{-2em}
\caption{A capability type-and-effect system for our actor language. A standard exchange rule is omitted.
Behaviour types are \emph{not} splittable.
}
\label{fig:types}
\end{figure}
The language in the previous section permits actors to send messages to other actors that they are not prepared to handle --- messages of types that may not be dealt with by the target actor's behaviour at the time of message delivery.
This section develops a type-and-effect system to ensure such errors are detected statically by the type checker.
Conceptually, there are two aspects. The first aspect, which gives rise to the name of our technique, is to equip actor references with a capability structure that restricts how they are used. Section \ref{sec:actorcaps}
explains that structure and how the use is enforced. Section \ref{sec:effects} then explains how the system ensures consistency between an actor capability sent to another actor and the subsequent behaviours of a given actor.

Capabilities and effects are both formal languages over the set of types.

\subsection{Substructural Actor Capabilities and Splitting}
\label{sec:actorcaps}

Actor capabilities in our type system are equipped with a \emph{formal language} $L$ drawn from $\mathsf{Type}^*$, 
and thus describe the sequence of message types that may be sent via that actor reference.\footnote{
Technically the set of all types forms an infinite alphabet (one may have, at a minimum, $n$-ary tuples of unit for each $n\in\mathbf{N}$),
but in the absence of polymorphic recursion, any given program only mentions a finite number of types (those used in behaviour case labels),
and we assume $\mathsf{Type}$ denotes that set, avoiding subtleties of infinite alphabets.}
Because the language describes both a restriction on the use of the value and a guarantee to the possessor that uses within that
restriction should be valid, we call these actor references capabilities.
An invariant we maintain is that the combined use of any and all actor capabilities (references) 
to some actor, in any interleaved order, should be allowed and handled by the actor they refer to --- 
after all, if each capability is held by a different actor, they could potentially be used in any order!
Such interleaving can be represented by the shuffle ($\shuffle$) operator.
The shuffle of two words
$x=x_1\ldots x_k$ and $y=y_1\ldots y_k$ (with $x_i$ and $y_i$ being possibly-empty subwords, not necessarily characters)
 is defined as
\[
    x_1\ldots x_k\shuffle y_1\ldots y_k = x_1y_1\ldots x_ky_k
\] 
which is lifted to languages as
\[
    L_1 \shuffle L_2 = \{ x\shuffle y \mid  x\in L_2, y\in L_2 \}
\]
That is, the shuffle of two languages consists of all possible interleavings of words from each language.

This gives us a way to control duplication of capabilities, to share actor references without increasing the permissions available
to some code. Consider the case of an actor reference with type $\mathsf{ActorRef}(\mathsf{nop}^*\cdot\mathsf{act}\cdot\mathsf{nop}^*)$
(momentarily assuming our languages are described by regular expressions, which is not a requirement of the formal development).
This reference can be used to send as many \textsf{nop} messages as desired or necessary, but may only be used to send an \textsf{act}
message once --- exactly once. Simply duplicating that reference at the same type results in \emph{two} actor references which could \emph{each}
be used to send an \textsf{act} message, and repeated duplication would then permit sending unlimited such messages.
Instead, we borrow an idea from the reference capability literature, of \emph{splitting} types with such restrictions~\cite{gordon2013rely,gordon2017verifying,militao2014rely,militao2016composing,CastegrenW16,CastegrenW17}.
We borrow Gordon's notation~\cite{gordon2013rely,gordon2017verifying} in Figure \ref{fig:types}:
 $\Gamma\vdash \tau \prec \tau' \divideontimes\tau''$ justifies splitting a variable of type $\tau$ into two copies, one of type $\tau'$ and the other of type $\tau''$.
This is used by the contraction rule (\textsc{T-Contract}). We call primitive types and closures \emph{reflexively splittable}
because they may be split without change (truly copied). Pairs split element-wise (\textsc{S-Pair}).
Actor references split using language shuffle (\textsc{S-ActorRef}).
Thus we could split our example actor reference into one of type $\mathsf{ActorRef}(\mathsf{act})$ and one of type $\mathsf{ActorRef}(\mathsf{nop}^*)$,
because the shuffle of those two protocol languages is contained in the original:
\[
    \inferrule*[right=S-ActorRef]{
        \mathsf{act}\shuffle\mathsf{nop}^*\subseteq\mathsf{nop}^*\cdot\mathsf{act}\cdot\mathsf{nop}^*
    }{
        \Gamma\vdash \mathsf{ActorRef}(\mathsf{nop}^*\cdot\mathsf{act}\cdot\mathsf{nop}^*) \prec 
            \mathsf{ActorRef}(\mathsf{act})
            \divideontimes
            \mathsf{ActorRef}(\mathsf{nop}^*)
    }
\]

Note that behaviours are \emph{not} splittable --- they are linear, and the absence of a rule for splitting behaviours
is by design. This allows a behaviour to capture a non-reflexively-splittable variable and use it.
\emph{Recursive} behaviours are encoded via recursive lambda expressions which return behaviours,
and lambdas may only capture reflexively-splittable values --- so any substructural
values used in such behaviours are freshly created each time the recursive function is invoked.

\paragraph{Consumption}
The point of use for an actor reference is via sending a message (\textsc{T-Send}).
This rule types the message to send, and uses an \emph{updating} typing judgment on paths to both retrieve the type of the path prior to use
(a variable, or ``field chasing'' from a variable bound to a tuple), and update that type (either updating the type of the variable directly,
or updating the type of appropriate --- possibly nested --- tuple components) to reflect sending a message of type $\tau$.
The judgment returns the updated type environment. This structure is chosen because our type system uses contraction deliberately,
and may have multiple bindings for the same variable; separately typing the path and updating the output type environment would introduce difficulties
ensuring the binding used for the pure typing was the same variable updated after the send, and it is simpler to do both at once.

$\tau^{-1}L$ is the derivative operator on formal languages, which includes all possible \emph{completions} of words in $L$
that start with the chosen element of the alphabet:
\[ \alpha^{-1}L = \{ w \mid \alpha\cdot w\in L \} \]

Of course the purpose of our actor reference type is two-fold: to enforce a usage discipline on code that holds an actor reference (addressed above), and to
guarantee that uses according to that discipline are valid. The second purpose is partly addressed by splitting
(which ensures an actor cannot increase the set of message sequences it can send to another actor)
and partly addressed by effects (which allow relating the capabilities that exist for an actor to its intended behaviours).

\paragraph{Program Typing}
If actors only run in response to messages, how does the system begin? 
A program is given by the behaviour $e$ for an initial actor, which must accept (at least) the unit type as a message.
The initial configuration for a system has
actor pool $\{\ell_0\mapsto e\}$ and global message queue $\{\ell_0,\ell_0\mapsto((),\mathbf{0})::\mathbf{nil}\}$, 
from which point the only available reduction (which \emph{is} available) is to deliver the unit message to the initial actor. So top-level programs are typed by
\[
\inferrule*[left=T-Program]{
    \emptyset\vdash e : \mathsf{Beh}(\overline{\tau})\\
    \mathbf{0}\in\overline{\tau}
}{
    \vdash e \; \mathsf{prog}
}
\]

\subsection{Consistency of Actor Capabilities, Behaviours and Effects}
\label{sec:effects}
To discuss consistency between behaviours and capabilities, we must discuss the typing of behaviours.
The type $\mathsf{Beh}(L)$ characterizes a behaviour which will accept message sequences in $L$,
though this does not fully characterize the future behaviour of the actor using this behaviour.
$L$ is intuitively an over-approximation of how all \emph{existing} capabilities to send to that behaviour could collectively
be used --- all capabilities that existed \emph{prior} to that behaviour handling a message. Note that this
does \emph{not} require knowing the exact history of the program: it is limited to the actor using the behaviour,
and refers only to the \emph{unused} capabilities, not past usage of capabilities. This information is necessary for
typing behaviours because they represent \emph{unresolved promises} for future message acceptance.
At the same time, an actor with such a behaviour installed may end up accepting more than $L$ in the future,
because when the behaviour handles a message, it may create additional obligations for its actor to handle
in the future, which we track using a static effect system.

\paragraph{The Effect of Capability Creation}
The starting point for all effects is \textsc{T-Self}, which captures the creation of a new
self-reference capability that is usable according to some particular language.
Other rules simply accumulate an over-approximation of created capabilities.
Subexpressions that run together in order (such as a function, argument, and latent function effect in an application expression per \textsc{T-App})
have their effects shuffled together with $\shuffle$, as collectively all created capabilities
could be used in an interleaved fashion.
Expressions that may run alternatively  are combined
with union --- the rule for conditionals combines the conditional effect $L_c$ with the
true- and false-branch effects $L_t$ and $L_f$ as $L_c\shuffle(L_t\cup L_f)$, capturing that the condition
always runs before either branch, but it is statically unknown which branch will execute.

Note that $\shuffle$ distributes over $\cup$ on both sides, so with $\{\epsilon\}$ as the unit for $\shuffle$,
these effects form a total commutative effect quantale~\cite{gordon2021polymorphic}. It follows both from that framework's soundness results
(and from our later proofs) that the effect of an expression over-approximates the shuffle-interleaving of all self-capabilities created during a local
expression's reduction. 

Note that the system tracks the \emph{creation} of new capabilities for the current actor, \emph{not} which capabilities are \emph{actually sent} to other
actors! This is both simpler than distinguishing between sending a self-reference or another actor's reference, and also reflects the fact
that an actor can send itself messages, in which case those would need to be handled by later behaviours.\footnote{It is of course also conservative:
an actor may create a self-reference that it then discards, but the discarded capability will still be captured in the effect
and become a new (un-dischargeable) obligation for the actor.}

\paragraph{Typing Behaviours}
\textsc{T-Beh} in Figure \ref{fig:types} is subtle, but captures the relationship between past promises, locally-created capabilities,
and future message acceptance.
Unsurprisingly, the rule requires typing each branch of the behaviour's body,
with the expectation that the result is a new behaviour (which will be used to handle future messages).
Those messages the actor handles in the future may come from one of two sources, each reflected in an argument to the
shuffle ($\shuffle$) operator.
The left argument $\tau_i^{-1}L$ models the fact that each branch of the behaviour handles a different incoming message
type (which must have been justified by a previously-existing capability accounted for by $L$),
but any remaining capabilities that existed before might still exist and be usable to send further messages.
So the left argument captures that the obligation to accept an initial $\tau$ has been discharged.
The right argument $L_i$ corresponds to new capabilities to the current actor which this message-handler may have
created, which are then \emph{additional} obligations for the future which did not previously exist,
so are not reflected in the original $L$. $L_i$ is guaranteed to include all new obligations because it is also
the \emph{effect} of the message handler.
So requiring the behaviour returned after handling $\tau_i$ to have type $\mathsf{Beh}\left((\tau_i^{-1}L)\shuffle L_i'\right)$
ensures that both previously-existing obligations and newly-created obligations will be supported by the next handler.

For example, in the \textsf{nop}/\textsf{act} example of the previous section,
if the actor has only given out that one capability discussed previously,
$L$ would be
$\mathsf{nop}^*\cdot\mathsf{act}\cdot\mathsf{nop}^*$.
In typing the handler for \textsf{act}, the derivative of that history assumption by \textsf{act}
would be $\mathsf{nop}^*$, reflecting that the one promised handling of \textsf{act} was settled,
and if that was the only such capability distributed, then no future behaviour would necessarily be forced
by that old capability (which was now partly consumed) to handle additional \textsf{act} messages. The derivative handles this.
However, if in response to the \textsf{act} message the handler created a new capability for the actor which granted permission
to send another \textsf{act} message, then the future behaviour would need to reflect that, hence the requirement
that the next behaviour also handle the languages of all newly-created capabilities --- incorporated via $\shuffle$.

\paragraph{Spawning New Actors, and Initial Capabilities}
The one remaining matter to attend to is creating an actor with a specific behaviour --- 
the rules discussed so far ensure that \emph{if the history assumptions are accurate they remain accurate},
and \textsc{T-Spawn} establishes the initial accuracy.
\textsc{T-Spawn} requires that the assumed protocol $L'$ for the initial external capability 
is a subset of the language $L$ that the spawned behaviour assumes exists prior to its execution.

\subsection{Soundness}
\label{sec:soundness}
\ifExtended
Because the safety property we are interested in --- that an actor never receives a message it does not expect ---
is one whose violation results in stuck reduction (the $\tau\in\overline{\tau}$ antecedent of \textsc{E-MsgHandled} would not
be satisfied), syntactic methods are adequate to show that the type system ensures our desired property.

\begin{figure}[t]\scriptsize
\begin{mathpar}
\inferrule*[left=\scriptsize T-LocSplit]{
}{
    \Sigma,\ell\mapsto L,\Sigma';\Gamma\vdash \ell : \mathsf{ActorRef}(L) \dashv \Sigma,\Sigma';\Gamma\mid \{\epsilon\}
}
\and
\inferrule*[left=\scriptsize T-LocPath]{
    \Sigma;\Gamma\vdash \ell : \mathsf{ActorRef}(L) \dashv \Sigma';\Gamma \mid \{\epsilon\} \\
    \tau^{-1}L\neq\emptyset
}{
    \Sigma;\Gamma\vdash_\mathsf{path} \ell : \mathsf{ActorRef}(L) \leadsto \mathsf{ActorRef}(\tau^{-1}L)\dashv \Sigma'\uplus[\ell\mapsto \tau^{-1}L] ; \Gamma
}
\and
\inferrule*[left=\scriptsize T-LocContract]{
    L' \shuffle L''\subseteq L\\
    \Sigma,\ell\mapsto L',\ell\mapsto L'',\Sigma';\Gamma\vdash e : \tau \dashv \Sigma'';\Gamma' \mid \chi
}{
    \Sigma,\ell\mapsto L,\Sigma';\Gamma\vdash e : \tau \dashv \Sigma'';\Gamma' \mid \chi
}
\and
    \inferrule*[left=\scriptsize T-Lam]{
        \Gamma\vdash \Gamma \prec \Gamma\divideontimes\Gamma\\
        \Gamma \vdash \Sigma \prec \Sigma\divideontimes\Sigma\\
        \Sigma;\Gamma,f:\tau\xrightarrow{\chi}\tau',x:\tau\vdash e : \tau'\dashv\Sigma,\Sigma';\Gamma\mid\chi
    }{
        \Sigma;\Gamma\vdash (\lambda_f x\ldotp e) : \tau\xrightarrow{\chi}\tau'\dashv \Sigma;\Gamma\mid\{\epsilon\}
    } 
\end{mathpar}
\caption{Local Runtime Typing}
\label{fig:runtime_typing}
\end{figure}

As is typical for syntactic methods in the presence of resources, we require a runtime type environment called a
\textsf{Local Capability Summary} for dynamically
allocated values (actors), analogous to the type environment.
To simplify the parity with type environments under substitution, 
we use multimaps $\Sigma \in \mathsf{Location}\rightharpoonup\mathcal{P}(\mathcal{L}(\mathsf{Type}^*))$ (mapping actor locations
to the languages they have already committed to accept). 
We add this to the typing derivation (Figure \ref{fig:runtime_typing}).
Notice that \textsc{T-LocPath} interacts with \textsc{T-LocSplit}. \textsc{T-LocPath}'s location typing antecedent
can only be proven using \textsc{T-LocSplit} precisely once (possibly following one or more uses of \textsc{T-LocContract}).
This means that the antecedent deals with adequate splitting of permissions to $\ell$, removes appropriate permission from $\Sigma$,
and allows the path typing to put the (updated) capability back.
Just as in the source-level type system, capabilities for the resulting value are simply trusted (not reflected in output $\Sigma$ or $\Gamma$).
Also notice the similarity between how \textsc{S-ActorRef} interacts with \textsc{T-Contract}, and \textsc{T-LocContract}.

We establish consistency between an actor environment, message queues, and dynamic actor environment via another relation
(treating behaviour values labeled with their language as pairs of the value and language for convenience):
\[
\inferrule*[left=T-LocalState]{
    \forall (\ell\mapsto v_b,L_b)\in\sigma\ldotp
        \Sigma;\emptyset\vdash v_b : \mathsf{Beh}(L_b) \dashv \emptyset \mid \{\epsilon\}
        \land L \le_\shuffle\mu(\ell)^{-1}L_b
}{
    \vdash \sigma;\mu : \Sigma
}
\]
Local state typing is subtle: it is used only with \emph{local} actor pools, of actors spawned during local reduction.
Hence every actor in $\sigma$ must be covered by $\Sigma$, but $\Sigma$ may mention additional actor locations.
$\mu$ is the local queue of outgoing messages \emph{from
the actor being type-checked}.

The form of type preservation is then largely familiar:
\begin{lemma}[Local Type Preservation]
\label{lem:local_type_preservation}
If $\vdash \sigma;\mu: \Sigma$ and \mbox{$\Sigma;\emptyset\vdash e : \tau \dashv \Sigma'';\emptyset \mid L$} and
$\mu;\sigma;e \xrightarrow{L'}_\ell \mu';\sigma';e'$, then
there exists a $\Sigma'$, $L''$, $\overline{\Sigma_{\textsf{sfx}}}$, and $\mu_\textsf{sfx}$ such that
\begin{enumerate}
\item \mbox{$\Sigma';\emptyset \vdash e' : \tau \dashv \Sigma'';\emptyset\mid L''$}
\item $\vdash \sigma';\mu' : \Sigma'$
\item $L'\shuffle L'' \subseteq L$ \label{goal:effect_rel}
\item $\mu\oplus\mu_\textsf{sfx}=\mu'$ (i.e., $\mu$ is a prefix of $\mu'$)
\item \label{goal:other_conservation} For all $\ell'\in\mathsf{dom}(\Sigma)$, $\ell'\neq\ell\Rightarrow \Sigma'(\ell')\shuffle\mathsf{caps}(\ell',\overline{\Sigma_\textsf{sfx}}) = (\mu_\textsf{sfx}|_{\ell'})^{-1}\Sigma(\ell')$
\item \label{goal:self_conservation} $\Sigma'(\ell)\shuffle\mathsf{caps}(\ell',\overline{\Sigma_\textsf{sfx}}) = \left((\mu_\textsf{sfx}|_{\ell})^{-1}\Sigma(\ell)\right)\shuffle L'$
\item $|\mu_\textsf{sfx}|_{\ell}| = |\Sigma_\textsf{sfx}|$ and $\forall \langle(v,\tau_v,\_),\Sigma_v\rangle \in \langle\mu_\textsf{sfx}|_{\ell},\overline{\Sigma_\textsf{sfx}}\rangle\ldotp \Sigma_v;\emptyset\vdash v : \tau_v \dashv \emptyset;\emptyset\mid\{\epsilon\}$
\end{enumerate}
\end{lemma}

\paragraph{Subtleties of Preservation}
While much of this follows typical forms for preservation proofs --- ensuring that after reduction
there is a new state type matching the new state and the simplified term has the same result type --- we
will elaborate on some differences. The handling of effects is standard for a sequential effect system~\cite{gordon2021polymorphic}:
the reduction step yields a dynamic effect for that reduction ($L'$), and after reduction the new expression may have a different
effect due to having fewer effectful operations left ($L''$). But these are all related (outcome \ref{goal:effect_rel}) --- the observed effect sequenced with
the effect of the remaining expression must be less than the original effect, for the original effect to be a sound bound on behaviour.

The less typical portion deals with message queues and the state type itself.
In languages with a heap type, it is common to require the new state type $\Sigma'$ to simply extend $\Sigma$ with
additional entries --- but this is because the type of a heap cell is typically invariant.
Our restriction on the new actor pool permits the addition of new entries (for newly-created actors), but also
permits updates to the capabilities assumed for existing actors, and as a multimap permits
multiple entries per actor. Fortunately, the suffix $\mu_\textsf{sfx}|_{\ell'}$ of messages newly-sent to
the actor at $\ell'$ exactly captures the usage this reduction has made of actor capabilities for $\ell'$, so the update for those
entries may be computed by taking the derivative of the initial $\Sigma(\ell')$ by the new messages.
The one exception is when $\ell'$ is the location $\ell$ of the running actor. In this case, we must also handle
the reduction dynamically creating new capabilities/obligations.

At least, that is the intuition, but this becomes slightly more complex in two ways.
First, our notation hides an important subtlety because these capability environments
are multimaps, not simply maps. We define $\Sigma(\ell)={}^{~\shuffle}_{l\mapsto L}L$,
that is, the difference between the \emph{shuffled collection of all capabilities} to a certain actor before and after evaluation is related
by the derivative.
Second, items \ref{goal:other_conservation} and \ref{goal:self_conservation} must deal with not only
the use of capabilities by the actor, but also the \emph{transfer} of capabilities by sending outgoing messages.
For example, if an actor's handler for a particular message has the protocol $\tau_1\cdot\tau_2\cdot\tau_3$, the actor might
use it to send $\tau_1$, and then send the capability for the remaining $\tau_2\cdot\tau_3$ to another actor.

That is, whatever is left of capabilities for a specific actor after use for message sends during reduction ($(\mu_\textsf{sfx}|_{\ell'})^{-1}\Sigma(\ell')$) must
be equal to the (shuffle) combination of the remaining capabilities as seen by the type judgment ($\Sigma'(\ell')$) and those implicitly embodied
in the outgoing messages. 
This is dealt with via $\mathsf{caps}(\ell',\overline{\Sigma_\textsf{sfx}})$, which is defined as $\shuffle_{\Sigma\in\Sigma_{\textsf{sfx}}}(\Sigma(\ell'))$.
We use only the new suffix for this, because the capabilities embodied in messages in the initial queue $\mu$ are not part of $\Sigma$.
We call these properties (\ref{goal:other_conservation} and \ref{goal:self_conservation}) \emph{capability conservation} to emphasize that
capabilities are not lost.\footnote{Though this is technically a slight misnomer, as unlike the laws of thermodynamics this name
alludes to, capabilities can genuinely be used up and exhausted --- but they don't just disappear.}

\paragraph{Substitution}
Due to the substructural nature of our type system, typing after substitution is sensitive to the number of occurrences of the replaced
variable. Because our contraction rule may give each split version of a variable a \emph{different} type, we require $n$ typing derivations
of the value for $n$ bindings of the replaced variable. The actual use cases for this lemma simply replace a single variable binding
(when substituting actual arguments in place of formal parameters), but the generalization is required to establish the proof.

\begin{lemma}[Substitution Preserves Typing]
\label{lem:subst_pres_type}
If \mbox{$\Sigma;\Gamma,\overline{x:\tau'}^n\vdash e : \tau \dashv \Sigma';\Gamma'\mid\chi$} and \mbox{$\overline{\hat{\Sigma};\emptyset\vdash v : \tau' \dashv \hat{\Sigma'};\emptyset\mid\{\epsilon\}}^n$}
and $x\not\in\Gamma$,
then there exists a $\hat{\Sigma'}$ such that 
\[\Sigma,\overline{\hat{\Sigma}}^n;\Gamma\vdash e[x/v] : \tau \dashv \Sigma',\overline{\hat{\Sigma'}}^n;\Gamma'\setminus x \mid \chi\]
\end{lemma}
Here $\tau'$, $\hat{\Sigma}$, and $\hat{\Sigma'}$ are indexed up through $n$ occurrences, while $v$ and $x$ are fixed choices of variable name and
value. Substitution deals with $n$ separate value typings because there may be $n$ occurrences of $x$ in a given type environment,
all replaced at once.
\begin{proof}
By induction on the typing of $e$.
Most cases are straightforward. The critical cases are the contraction rule \textsc{T-Contract} and the variable rule \textsc{T-Var}.
Intuitively, every use of \textsc{T-Contract} should be replaced by a use of \textsc{T-LocContract} for each location occurring in $v$,
and \textsc{T-Var} should be replaced by a use of \textsc{T-LocSplit}.
\begin{itemize}
\item Case \textsc{T-Contract}: In the case where the contracted variable is not $x$ the result follows from the inductive hypothesis.
      On the other hand, if the contracted variable \emph{is} $x$, then it must be the case that $\Gamma_0$ (this is inside a derivation,
      and not the top-level $\Gamma$) justifies splitting $\tau$ into $\tau'$ and $\tau''$. This split corresponds to checking, for each position in
      $\tau$ (and $\tau'$ and $\tau''$, which have the same shape) that contains an actor reference, that the language (say $L$) decomposes into
      a shuffle of two languages (let's say $L'$ and $L''$). By induction, if $\ell$ is in the corresponding position in $v$,
      then \textsc{T-LocSplit} must find an entry $(\ell\mapsto L)\in\hat{\Sigma}$, for the same $L$. This language then decomposes also into $L'$ and $L''$.
      We can repeat this reasoning for every position in $v$ containing a location, yielding a $\hat{\Sigma_2}$ and $\hat{\Sigma_3}$ which each validate
      typing the value at types $\tau'$ and $\tau''$ respectively.
      Of course, this applies to only one of the value typings, corresponding to the particular choice of binding for $x$. If the $i$-th binding of $x$ in the
      type environment is split, then the $i$-th value typing among our hypotheses is split accordingly. Then the result follows from the inductive hypothesis
      applied to the now $n+1$ bindings of $x$ in the antecedent of the contraction rule and the now $n+1$ typings of $v$.
      The choice of $\hat{\Sigma'}$ follows from the inductive application.
\item Case \textsc{T-Var}: If the variable is not $x$ this case is trivial. If this is typing the variable being replaced,
      then the typing of $x$ used is one of the $n$ in our assumptions, meaning we can select the corresponding $i$-th value typing.
      We then set $\hat{\Sigma'}$ to $\overline{\hat{\Sigma}}^{1..i-1},\overline{\hat{\Sigma}}^{i+1..n}$ and can derive the appropriate typing
      from the $i$-th typing itself and weakening.
\end{itemize}
\end{proof}
The above proof, as given, relies on a close correspondence between positions of actor reference types in types and positions of locations in values.

\paragraph{Proof of Local Type Preservation}
Now we are ready to prove local type preservation.
\begin{proof}[Lemma \ref{lem:local_type_preservation}]
By induction on the reduction relation, followed by inversion in each case on the assumed typing derivation.
The only interesting cases are:
\begin{itemize}
\item Case \textsc{E-Spawn}: In this case $\Sigma'$ extends $\Sigma$ with the new actor's behaviour language.
\item Case \textsc{E-Send}: In this case, inversion on the destination location typing (via \textsc{T-Loc})
ensures that the type assumed for the send not only permits sending the message type first, but that this is consistent with $\Sigma$.
Expression typing is straightforward. We must set $\Sigma''=\Sigma'[\ell_\mathsf{dst}\mapsto\tau^{-1}\Sigma'(\ell)]$ since
a message of type $\tau$ was enqueued for $\ell_\mathsf{dst}$.
We are sending a value $v$, which by assumption in this case is well typed under $\Sigma;\emptyset\vdash v : \tau_\textsf{msg} \dashv \Sigma';\emptyset \mid \{\epsilon\}$.
Because value typing does not \emph{change} capabilities, only consumes them outright, let $\Sigma_\textsf{sfx}$ be $\Sigma'\setminus\Sigma$ ---
the capabilities consumed in typing $v$. This can then be used to type the value in isolation (\mbox{$\Sigma_\textsf{sfx};\emptyset\vdash v : \tau_\textsf{msg} \dashv \emptyset;\emptyset\mid\{\epsilon\}$}).
\item Case \textsc{E-Self}: In this case the annotated language $L_c$ for the new capability is raised as an effect originally,
and this would be the $L'$ raised by the semantics.
The resulting value (per \textsc{T-Loc}) now has effect $\{\epsilon\}$; letting that be $L''$, we have $L_c\shuffle\{\epsilon\}\subseteq L_c$.
To preserve state typing, $\Sigma'$ must update actor $\ell$'s entry to $\Sigma(\ell)\shuffle L_c$.
\item Case \textsc{E-App1}: This rule (omitted from Figure \ref{fig:lang}) reduces the function position of an application, stepping $(e_f\;e_a)$ to
$(e_f'\;e_a)$ when $e_f$ reduces in one step to $e_f'$ (propagating actor creation and message queues appropriately), and is emblematic of other
``contextual'' rules. The inductive hypothesis, applied to the reduction from $e_f$ to $e_f'$ gives the new $\Sigma'$ directly, satisfying its constraints,
and what remains is to ensure the effects are valid when constructing the post-reduction typing derivation (the type aspects are standard).
In particular, the inductive hypothesis gives an $L_{f'}$ and says that $L'\shuffle L_{f'}\subseteq L_f$
(where $L_f$ was the static effect of $e_f$ prior to reduction). Assuming the inversion on the pre-reduction application gave an effect $L_a$ for the argument
and $L_{\textsf{latent}}$ as the latent effect of the function being applied, the original effect (before reduction) was $L_f\shuffle L_a\shuffle L_{\textsf{latent}}$.
We require an $L''$ which is an effect of the reduced application expression $(e_f'\;e_a)$, satisfying $L'\shuffle L'' \subseteq L_f\shuffle L_a\shuffle L_{\textsf{latent}}$.
Choosing $L''=L_f'\shuffle L_a\shuffle L_{\textsf{latent}}$ satisfies this constraint, using the constraint from the inductive hypothesis.
\end{itemize}
Other cases either follow immediately from the inductive hypothesis and standard type substitution lemmas, are trivial (e.g., arithmetic,
pairs), or are subexpression reductions similar to the shown \textsc{E-App1}.
\end{proof}
A careful reader may notice that for a single step, at most one message can be sent, at most one actor can be created, and at most one capability can be created,
and these are mutually exclusive --- so the restrictions on $\Sigma'$ could be stricter. However, the general form used above is both
sufficiently precise for the lemma to go through, and also the \emph{same form as for multi-step local preservation}. Therefore we omit details of the multi-step local
preservation, which is defined identically to Lemma \ref{lem:local_type_preservation} except for using the multi-step relation.
The use of the derivative with respect to newly-sent messages helps to preserve state typing even with multiple messages sent,
as derivatives satisfy $(ww')^{-1}L=w'^{-1}w^{-1}L$ (e.g., taking $w'$ to come from a suffix of the transitive reduction). 

Global preservation relies on a slightly different state typing that handles global message queues, plus a global map of local capability summaries
$\Pi \in \mathsf{Location} \rightharpoonup \textsf{Local Capability Summary}$ summarizing each actor's assumed capabilities.
\[
\inferrule*[left=T-GlobalState]{
    \forall \ell\ldotp \ell\in\mathsf{dom}(\Pi) \Rightarrow (\ell\in\mathsf{dom}(\sigma)\land \forall \ell'\in\mathsf{dom}(\Pi(\ell))\ldotp \ell'\in\mathsf{dom}(\sigma))\\
    \forall (\ell\mapsto v_b,L_b)\in\sigma\ldotp \left(\shuffle_{\scriptsize \ell'\in\mathsf{dom}(\sigma)}\Pi(\ell')(\ell)\le_\shuffle \gamma(\ell)^{-1}L_b\right)\\
    \forall (\ell\mapsto v_b,L_b)\in\sigma\ldotp 
	\Pi(\ell);\emptyset\vdash v_b : \mathsf{Beh}(L_b) \dashv \emptyset \mid \{\epsilon\}
}{
    \vdash \sigma;\gamma : \Pi
}
\]
This global state invariant both captures the consistency across all actors, modularly.
Unlike local typing, it deals with all actors in the system, strictly between message deliveries (as opposed to local typing
only checking consistency with newly-allocated actors mid-message delivery). This simplifies reasoning about the ordering
of message deliveries. When an actor at $\ell$ receives a message of type $\tau$, the behaviour handler for $\tau$
must be run assuming that the relevant $\tau$ is \emph{no longer} possible to send (e.g., if the actor's behaviour was
prepared for only one $\tau$ ever, such as an initialization message). Yet while that handler is executing we cannot
yet know the future behaviour for the current actor. So there is no consistent pair of global states and capability types
\emph{during} execution, because the behaviour for after delivery of $\tau$ is still being computed.

Another subtlety of \textsc{T-GlobalState} is the meaning of $\gamma(\ell)$. This still yields a \emph{single}
sequence, which is an unspecified shuffle of the message sequences in flight \emph{to} $\ell$ from all other actors.

\fi

\section{Related Work}
\label{sec:relwork}
Our system draws on many areas of prior work, including substructural type systems and reference
capabilities, but those were discussed earlier.
We focus this section not on an exhaustive explanation of the many sources of inspiration
for this work (largely covered by related work discussions of the previously-cited papers),
but instead on comparing our work against other approaches to ensuring that actors (or active objects) and
processes handle specific message sequences and send compatible message sequences to others.

A classic point of reference for this is Nierstrasz's work~\cite{nierstrasz1993regular} on \emph{regular types},
assigning finite-state-based specifications of object behaviour, as the sequence of messages accepted and the resultant
responses. His exposition explores the structure of these specifications and informally how they might apply
to objects, but provides no application of these types to a concrete programming language, and focuses on composing
applications by way of one object using another as its sole client.
This focus on the behaviour of the active object itself over time is also nearly dual to ours: regular types
for active objects describe the process behaviour summatively, while in our (source-level) type system, no
summative language-oriented description of an actor's accepted messages exists! Instead static knowledge
of an actor's accepted language can only be assembled from various capabilities.
Active objects, as nicely captured in de Koster et al.'s taxonomy of actor language variants~\cite{de201643},
do not permit changing the set of accepted messages over time. Support for such changes is an explicit goal of this work.
\looseness=-1

Our calculus is pure, but many popular actor frameworks are in languages with mutable state, and permit sharing
references across actors on the same machine~\cite{charousset2014caf,haller2012integration}.
The introduction of direct mutable heaps raises classic issues of balancing safety and efficiency of sending messages. 
Deep copying of state is always sound, but highly inefficient, especially when the data being sent is either immutable
or the sender discards its own access to the mutable data after sending.
This has motivated a great deal of work on various flavors of uniqueness and mutation control,
including a great deal of work applying the reference capability notions discussed in the introduction 
to track forms of unique ownership transfer or object immutability~\cite{gordon2012uniqueness,gordon2017verifying,CastegrenW16,CastegrenW17,clebsch2015deny,servetto2013balloon,giannini2019flexible}. Some of this work~\cite{gordon2012uniqueness,clebsch2015deny} has specifically been applied to actor systems.\footnote{Now that the relevant NDAs are long-expired, we can finally say in writing that the context for \cite{gordon2012uniqueness} was the Midori~\cite{duffyblog} derivative of Singularity~\cite{Fahndrich2006language,Hunt2007singularity}, where processes were essentially actors.}
However, incorporating two orthogonal flavors of capabilities --- the actor capabilities proposed here, 
along with the reference capabilities from existing work --- would distract from the main technical development of capabilities
for controlling message ordering in actor systems.
\looseness=-1

An alternative approach to managing state sharing and exchange between actors is the introduction of
specially-shared capsules. De Koster et al.~\cite{de2012domains,de2015domains} introduce \emph{domains}, which
are shared-memory constructs which actors asynchronously request access to (akin to asynchronous mutex acquisition).
Similarly, Cheeseman et al.'s \emph{cowns}~\cite{cheeseman2023concurrency} (named for \emph{concurrent ownership})
are shared resources accessed by asynchronously-spawned computations that run only when all required cowns are available at once.
Some of the most closely related work to ours also falls into this group:
Caldwell et al.~\cite{caldwell2024programming} describe a specialized data exchange environment called a \emph{dataspace}, as well as an effect
system describing behaviour over time for actors and data --- effectively dual to our approach of restricting actor behaviours with
capabilities. Similar to Skalka et al.~\cite{skalka2008types,Skalka2008}, their effects give abstract summaries of actor behaviour,
which can then be used to model-check properties of the program --- the effect types for both Skalka et al.'s work and Caldwell et al.'s work
give a model that is verified to soundly abstract the program code.
Establishing a formal relationship between our capabilities and behavioural effects like Caldwell et al.'s, akin to Craig et al.'s
formal relationship between flow-insensitive capabilities and flow-insensitive effects~\cite{craig2018} would be intriguing future work.

Session types~\cite{Honda2008,lagaillardie2020implementing}, choreographic programming~\cite{shen2023haschor,cruz2023formal,giallorenzo2024choral}, and multitier programming~\cite{weisenburger2020survey}
solve the issue of agreement between the messages accepted by a given process and message sent to it by others in the inverse way
of what we describe here. Instead of analyzing correctness of processes in isolation,
these approaches start with a single program describing all global behaviour, and a compiler generates
code (or in the case of session types, synchronization skeletons) for individual processes via a projection process.
These approaches are generally applied to \emph{synchronous} message passing, though 
Harvey et al.~\cite{harvey2021multiparty} and Neykova et al.~\cite{neykova2017multiparty} have adapted session types to actors (though
still using a global description of the full orchestration which is then projected for each of a fixed set of roles).
By contrast, our lack of a global type is a trade-off. On one hand, there is no complete closed description of behaviour separate from the code itself,
which can be useful for understanding the system as a whole.
On the other hand, actors in our approach can be specified and checked in isolation (e.g., as library components), without needing to know a priori how they fit into
a larger system, because any messages accepted by the actor beyond those described in the initial capability returned from \textsf{spawn}
are chosen by the actor itself.
Like Caldwell et al.'s work~\cite{caldwell2024programming}, the type of a process in these approaches describes \emph{all} of the process's future
behaviour, rather than only the promises it has made to its environment about its behaviour (as in our work).
\looseness=-1

Originating around the same time as classic binary session types~\cite{honda1998language}, Nielson and Nielson~\cite{nielson1993cml} described one of
the earliest flow-sensitive effect systems, for ensuring Concurrent ML (CML) processes sent and received \emph{synchronous} messages in an expected order.
This early similarity turned out to be quite relevant, as session types can be recast as an effect system (and vice versa)~\cite{orchard2016effects}.
Thus the idea of applying effect systems --- or an equivalent formalism --- to ensure correct message request ordering is itself far from new.
And the idea of using capabilities to restrict effects is also not new, going back past recent formal connections~\cite{craig2018}
to many earlier expositions of effects~\cite{henglein2005effect,Marino2009} in an implicit form.

Two aspects of our approach distinguish it from this established body of work.
First, this is to the best of our knowledge the first use of flow sensitive \emph{capabilities} (a form of coeffect~\cite{choudhury2020recovering})
to achieve safe ordering. While the relationship to a putative effect system achieving the same ends has not been formally pursued, and this
likely carries some of the downside of capabilities (vs. effects) outlined in our prior work~\cite{gordon2020designing} (which applied only to
flow-insensitive effect systems), it also carries a unique advantage in system design not discussed in that paper.
While effect systems may obtain greater precision in reasoning about side effects~\cite{gordon2020designing},
they do so at the cost of needing to ``reconstruct'' behaviour and check compatibility after the fact.
In contrast, our capability formulation of the problem inverts this checking, and places each actor (and its original allocation site)
in firm control of how other actors interact with it --- the protocols of the self-capabilities an actor sends to others
directly constrain those others' interactions a priori, localizing checks to a far greater extent than in CML or session types --- no direct global accounting of
behaviour is required to exist.

Second --- and a complementary factor in not requiring a global behavioural description to exist --- is our use of an effect system
to track capability \emph{creation}. This differs from most static capability work, which typically assumes all possible capabilities for something
exist upon its creation, and must be passed through the program. Instead we treat actors' retrieval of a new self-reference as a
\emph{dynamic} capability creation --- via effects. While it has long been recognized that both coeffects (e.g., capabilities) and effects are required
to reason about various combinations of independently-interesting static analyses and language semantics~\cite{gaboardi2016combining,dal2022relational}, ours is one of
very few examples of mixing coeffects and effects \emph{as part of a single static analysis}.

Another body of work with a more nuanced relationship to the work reported here is the concept of \emph{mailbox types},
as originally proposed by de'Liguoro and Padovani~\cite{de2018mailbox} and later implemented for an actor language with explicit mailboxes
by Fowler et al.~\cite{fowler2023special}. The core idea there is to drastically restrict aliasing of mailbox references,
and use the near-linearity to reason about the contents of the \emph{current actor's} mailbox, using commutative regular expressions.
Beyond their implementation of their system, and experience applying it to some benchmark programs, there are three key differences between this idea and actor capabilities.
First is that the behaviour tracking is almost inverted:
rather than restricting senders to only send what the recipient expects, the recipient is instead restricted to only \emph{request}
things from the mailbox which are known to have been sent to it. Mailbox types are predicated on the \emph{selective receive} operation
present in some actor frameworks, which retrieves a message of specified type. Thus mailbox types prevent an actor from issuing a selective
receive which would fail.
Second is that only message multiplicity is tracked, because of the commutative regular expressions; the use of selective receive
makes this the appropriate modeling choice, though most actor frameworks include both normal and selective receive, making
integration of our ideas and theirs interesting future work.
Third, they use a combination of second-class values and a mild relaxation of linearity to deal with aliasing of mailboxes and strong updates to mailbox types.
By contrast, we permit many aliases to each actor whose usage restrictions we track, by ensuring that each alias is independently usable (via $\shuffle$);
again, their choice is a reasonable one for their setting, as each mailbox is expected to be consumed by only one actor.

Program logics have also been used for reasoning about correctness of actor messages.
Gordon~\cite{gordon2019modal} proposed modal assertions of the form $@_l(P)$ inspired by hybrid logic~\cite{areces2007hybrid},
indicating that $P$ was true of the state of the actor at $l$. Thus assertions true at the sending actor
could be packaged in this way, and communicated as part of message invariants; a rely-guarantee-style relation constraining
state evolution worked together with a restriction to only send assertions stable with respect to that relation
ensured the assertions about other actors were never invalidated, but limited what could be proved to assertions over
monotonically-evolving portions of state (though CRDTs have shown this can be made more flexible than it initially appears~\cite{shapiro2011conflict}).
More recently some program logics (instantiations of \textsc{Iris}~\cite{jung2018iris}) have tied together
ideas from session types and verification for simple actor languages~\cite{hinrichsen2019actris,hinrichsen2024multris}. This work handles protocols similar to binary (for Actris)
or multiparty (for Multris) session types, but each message has not only a value but a set of assertions that the recipient may
assume upon receipt of the corresponding value.
This is much more powerful and flexible than Gordon's approach, at the cost of complexity. Gordon's system was prototyped as a library for the automated verification tool Dafny, %
while the modern approaches rely on \textsc{Iris} embedded in an interactive proof assistant.
The key point of comparison to the present work is that Actris and Multris do ensure the absence of unhandled messages,
but the specifications for doing so are essentially traditional session types. Compatibility when delegating
use of a channel to another process is checked via an ad hoc simulation search procedure akin to Milit\~ao et al.'s~\cite{militao2016composing}.
In contrast, our use of shuffle uncovers a clean algebraic structure underlying common uses of sharing and resharing
the addresses of other processes, though it is less general than Actris and Multris, where the correct continuation
of a protocol may depend on the specific value sent.

\begin{credits}
\subsubsection{\ackname}
Thanks to Gul Agha for his influential work on the actor model, which forms the basis for much of my research career (via the work
I am most known for~\cite{gordon2012uniqueness} and some of the work I am most fond of~\cite{gordon2019modal});
as well as some of what I most enjoy teaching (using actors when teaching distributed computing).
Agha's work~\cite{karmani2009actor,agha1986actors} has been the source for a significant amount of my own learning to think about concurrent programs
and their implementation, and by chance
he has had a direct influence on individual people who have influenced me during time in industry and academia~\cite{miller2005concurrency}.
Thanks also to Ivan Beschastnikh, for asking me in 2012 how I could apply reference capabilities to distributed systems. It took 13 years, but I think I’ve finally articulated a good answer,
even if handling faults remains future work.
Audience members at IWACO 2024 provided useful feedback on a more abstract version of the ideas presented here.
Thanks to Jonathan Schuster for bringing mailbox types to my attention, and to Lindsey Kuper for bringing my preprint to Jonathan's attention.
This work was supported by NSF Award \#CCF-2007582.

\end{credits}
\bibliographystyle{splncs04}

\end{document}